\documentclass[]{bytedance_seed}

\usepackage[toc, page, header]{appendix}
\usepackage{siunitx} 
\DeclareSIUnit{\kcal}{kcal}
\DeclareSIUnit{\GPUhour}{GPU\mbox{-}hour} 
\DeclareSIUnit{\e}{e}
\DeclareSIUnit{\bar}{bar}
\DeclareSIUnit{\angstrom}{\text{\AA}}
\DeclareSIUnit\Molar{\textsc{m}}

\usepackage{textgreek}
\usepackage{makecell}
\usepackage{float}
\usepackage{listings}

\title{THEMol dataset: Torsion, Hessian, and Energy of Molecules}

\author[1, \dagger]{Jiashu Liang}
\author[1, \dagger, *]{Tianze Zheng}
\author[1, \dagger]{Yu Xia}
\author[1]{Xingyuan Xu}
\author[1]{Xu Han}
\author[1]{Zhi Wang} 
\author[1]{Siyuan Liu} 
\author[1]{Ailun Wang}
\author[1]{Yu Liu}
\author[1, \S]{Shiqian Tan} 
\author[1, \ddagger]{Dongfei Liu} 
\author[1]{Zhichen Pu} 
\author[1]{Yuanheng Wang} 
\author[1]{Qiming Sun} 
\author[1, \S]{Xiaojie Wu} 
\author[1, *]{Wen Yan}

\affiliation[1]{ByteDance Seed}

\contribution[\dagger]{Equal contribution}
\contribution[*]{Corresponding authors}
\contribution[\ddagger]{Work done as intern at ByteDance}
\contribution[\S]{Work done at ByteDance}

\abstract{
    We present THEMol (Torsion, Hessian, Energy of Molecules), a massive open-source collection of quantum mechanical properties tailored for closed-shell organic molecules, with up to 50 heavy atoms.
    THEMol includes a Hessian subset with more than 3 million relaxed geometries with Hessian matrices, a TorsionScan subset with nearly 100 million constrained relaxed geometries with energies and forces, and relaxation-trajectory subsets (HessianRelax and TorsionScanRelax) that together comprise about 3 billion DFT calculations.
    The chemical space sampling is comprehensive, spanning twelve essential elements and diverse molecular architectures relevant to drug discovery, electrolytes, ionic liquids, and beyond.
    The dataset also features exhaustive conformational sampling through the TorsionScan and TorsionScanRelax subsets, including comprehensive in-ring and non-ring torsional scans.
    Furthermore, it contains an extensive library of Hessian matrices, computed at relaxed geometries, to capture critical second-derivative information of the potential energy landscape.
    Additionally, we supply electron density-derived atomic multipoles computed via the Minimal Basis Iterative Stockholder partition scheme.
    Organized into five distinct subsets (Hessian, TorsionScan, HessianRelax, TorsionScanRelax, and MBIS), the data encompasses optimized geometries, relaxation trajectories, and derived molecular properties.
    We anticipate that this massive and diverse dataset will significantly empower the development of highly accurate and transferable molecular potentials.
}

\date{\today}
\correspondence{Tianze Zheng at \email{zhengtianze@bytedance.com},  Wen Yan at \email{wen.yan@bytedance.com}}

\begin{document}
\maketitle

\section{Introduction}

High-quality quantum mechanical (QM) data for the potential energy surface of organic molecules are essential for parameterizing both molecular mechanics and machine learning (ML) potentials.
Large-scale QM datasets have laid the foundation for modern force field development.
Foundational datasets like QM9~\cite{ramakrishnanQuantumChemistryStructures2014} paved the way for small organic molecules, while the PubChemQC series scaled this approach to 221 million PM6-optimized geometries~\cite{nakataPubChemQCPM6Data2020} and 86 million DFT electronic structures~\cite{nakataPubChemQCB3LYP631G2023}.
More recently, the QCML dataset expanded the frontier with 33.5 million DFT and 14.7 billion semi-empirical calculations~\cite{ganscha2025qcml}.

Parallel to these efforts, transferable neural network potentials were catalyzed by the ANI and AIMNet dataset families.
ANI-1 introduced a dataset of 20 million off-equilibrium conformations~\cite{smithANI1DataSet2017} and demonstrated the first neural network potential achieving DFT accuracy at force field computational cost~\cite{smith2017ani1}.
ANI-1x extended this with DFT properties for diverse conformations~\cite{smith2020ani1x}.
Notably, it includes atomic charges, volumes, and multipoles computed from the Minimal Basis Iterative Stockholder (MBIS) partition scheme~\cite{verstraelen2016minimal}.
MBIS extracts these properties rigorously from a molecule's continuous electron density, enabling the development of vastly more accurate classical and machine learning force fields~\cite{pulido2024nonbonded}.
Later, ANI-1ccx achieved coupled-cluster accuracy through transfer learning~\cite{smithApproachingCoupledCluster2019}, while ANI-2x expanded the chemical space to include sulfur and halogens, covering seven elements in total~\cite{devereux2020ani2x}.
Most recently, ANI-1xBB introduced a reactive potential for small organic molecules~\cite{zhangANI1xBBANIBasedReactive2025}.
Concurrently, the AIMNet family demonstrated robust multitask prediction~\cite{zubatyuk2019aimnet, anstineAIMNet2NeuralNetwork2025}, eventually advancing generalized reaction modeling through AIMNet2-NSE and AIMNet2-rxn~\cite{kalitaAIMNet2NSETransferableReactive2026a, anstine2025aimnet2rxn}.

Recent dataset initiatives have further diversified chemical space coverage and targeted specific physical domains.
The Open Molecules 2025 (OMol25) dataset is composed of more than 100 million density functional theory (DFT) calculations, providing comprehensive coverage of molecular systems~\cite{levine2025omol25}.
This is complemented by the Open Polymers 2026 (OPoly26) dataset for polymeric materials~\cite{levine2025opoly26}.
The QCell dataset offers comprehensive quantum-mechanical data spanning diverse biomolecular fragments~\cite{kabyldaQCellComprehensiveQuantumMechanical2026b}.
For pharmaceutical applications, SPICE~\cite{eastman2023spice}, GEOM~\cite{axelrod2022geom}, and QMugs~\cite{isert2022qmugs} specifically address the parameterization needs of drug-like molecules.
For reactive chemistry, Transition1x provides specialized data for building reactive ML potentials~\cite{schreinerTransition1xDatasetBuilding2022}, and HORM offers a large-scale molecular Hessian database for optimizing reactive potentials~\cite{zhao2025horm}.
For noncovalent interactions, DES370K provides gold-standard dimer interaction energies~\cite{donchev2021des}, BFDb offers comprehensive biofragment data~\cite{burnsBioFragmentDatabaseBFDb2017}, and the NCI Atlas catalogs hydrogen bonding systems~\cite{rezacNonCovalentInteractionsAtlas2020a}.
For biomolecular systems, AIMD-Chig explores the conformational space of proteins with ab initio molecular dynamics~\cite{wangAIMDChigExploringConformational2023}.
For Hessian data, Hessian QM9~\cite{williamsHessianQM9Quantum2025} and VIBFREQ1295~\cite{zapatatrujilloVIBFREQ1295NewDatabase2022} provide vibrational frequency benchmarks.

To address unmet needs in force field training, we introduce a massive open-source DFT dataset accumulated over a span of nearly four years.
Centered on organic molecules, this dataset covers twelve essential elements (H, B, C, N, O, F, Si, P, S, Cl, Br, and I) and includes molecules with up to 50 heavy atoms.
Except for the MBIS subset, all DFT data are computed at the B3LYP-D3(BJ)/DZVP level of theory, specifically chosen to be compatible with similar datasets generated by the Open Force Field initiative.
This level of theory has been shown to be an efficient choice for intramolecular potential energy surface exploration~\cite{beharaBenchmarkDFT2024}.
The dataset distinguishes itself through three key advantages.
First, it provides comprehensive chemical space sampling, covering a broad chemical space relevant to drug discovery, electrolytes, ionic liquids, and beyond.
We demonstrated a comparison in our previous publication~\cite{zhengByteFF24}, based on a subset of THEMol.
The chemical space sampling is further enhanced by protonation state enumeration to systematically explore a broader range of charged molecular states.
Second, it features comprehensive intramolecular potential energy surface exploration using both unconstrained relaxation and torsion scans, enabling an almost exhaustive search of local minima on the potential energy surface.
Third, it incorporates extensive Hessian matrix data computed at relaxed geometries, providing critical second-derivative information for accurate force field development.

To provide maximum flexibility for force field developers, the dataset is systematically organized into five distinct subsets: Hessian, HessianRelax, TorsionScan, TorsionScanRelax, and MBIS.
Ultimately, this comprehensive collection establishes an unparalleled foundation not only for training high-fidelity force fields, but also for diverse applications benefiting from potential energy surface and electron density information of organic molecules.
In particular, we hope THEMol, which focuses on single-molecule intramolecular PES, provides complementary information to the popular OMol25 dataset~\cite{levine2025omol25}, which is a great source for diverse and highly accurate intermolecular PES.

\section{Method}
\subsection{Molecular Fragments Curation}
The curated dataset was primarily sourced from a publicly accessible molecular database: UniChem~\cite{chambers2013unichem}.
In addition, several widely used force field training datasets~\cite{LuSFE2021, ThermoMLArchivevdw2021}, compounds curated from published ligand discovery studies~\cite{Wang2015, Wang2017, Schindler2020}, and proprietary in-house collections were incorporated to enhance the practical applicability of the dataset.
Molecules were initially filtered based on physicochemical descriptors, including the number of aromatic rings, polar surface area (PSA), quantitative estimate of drug-likeness (QED), element types, and hybridization types.
Following selection, molecules were cleaved into fragments with fewer than 70 atoms using our in-house graph-expansion algorithm.~\cite{zhengByteFF24, jinsong2024molecular, roosOPLS3eExtendingForce2019}
This procedure is designed to preserve local chemical environments.
In brief, this fragmentation algorithm traverses each bond, angle, and non-ring torsion in a molecule.
It retains the relevant atoms and their conjugated partners, trims the rest, and caps the cleaved bonds.
To reduce potential sampling bias introduced by our fragmentation algorithm and to increase chemical space coverage, we additionally sampled a large number of fragment-like molecules from UniChem and bypassed the fragmentation pipeline.
Next, we enumerated protonation states with predicted $pK_a$ values between 0.0 and 14.0 using Epik 6.5.~\cite{shelley2007epik}
This range is intended to cover most protonation states expected in aqueous solution.
After this initial curation, approximately 4 million fragment-like molecules spanning multiple net charges were passed to the next step of the data generation pipeline.

\subsection{Conformation Generation}
We generated initial 3D conformations from SMILES strings using RDKit.
We then optimized them with the geomeTRIC optimizer~\cite{Wang2016} at the specified QM level.
The resulting optimization trajectories were archived to form the {HessianRelax} dataset. 
These relaxed conformations were then used for Hessian calculations, forming the {Hessian} dataset.
To ensure structural integrity, each conformer was screened to confirm that no bond scission or formation occurred during relaxation.
Furthermore, the stationary points were validated by verifying that all Hessian eigenvalues—excluding the six near-zero modes corresponding to translation and rotation—were positive, confirming the identification of true local minima.

From the relaxed conformations, we selected unique torsions for torsion scans.
Because of their distinct topological constraints, non-ring and in-ring torsions were treated with different protocols.
For non-ring torsions, the dihedral angles were rotated in 15$^\circ$ increments relative to the optimized geometry.
The resulting 24 initial frames were then subjected to constrained optimization using geomeTRIC.
In contrast, in-ring torsions were scanned using a sequential, frame-by-frame approach.
We applied early stopping if the relative conformational energy exceeded 20~\si{\kcal\per\mol}.
Such high-energy conformers tend to contribute little under Boltzmann weighting in typical applications.
The scanned geometries were filtered to ensure bond consistency.
The optimization trajectories and final constrained geometries were collected to form the {TorsionScanRelax} and {TorsionScan} subsets, respectively.

\subsection{Quantum Chemical Calculations and Quality Filters}
We employed the B3LYP-D3(BJ)/DZVP level~\cite{stephens1994ab,dunning1989gaussian} of theory for all calculations outside the MBIS subset.
This choice provides a practical balance between computational cost and accuracy for intramolecular PES of organic molecules.~\cite{beharaBenchmarkDFT2024}
Furthermore, it aligns seamlessly with the reference data utilized in widely adopted force field parameterization workflows~\cite{boothroyd2023openff}.

For the MBIS subset, the PBE0 functional~\cite{adamo1999toward} was paired with the def2-TZVPD basis set~\cite{rappoport2010property} for the vast majority of molecules.
The PBE0 functional was selected because it has been reported to be among the more robust methods for predicting electron densities.~\cite{medvedev2017density,liangGoldStandardChemicalDatabase2025a}
Consequently, it also performs well for density-derived properties such as molecular dipole moments.~\cite{hait2018accurate}
However, for iodine atoms, the DZVP basis set was employed instead of def2-TZVPD.
This substitution is necessary because the def2-TZVPD basis set employs an effective core potential (ECP) for iodine, which replaces core electrons with a parameterized potential.
The MBIS partition scheme~\cite{verstraelen2016minimal} fundamentally requires the complete all-electron density to rigorously extract atomic charges and multipoles, which DZVP provides.

To ensure the high quality of our dataset, we implemented the following systematic filtering steps:
\begin{itemize}
    \item \textbf{Spin State Filtering:} All open-shell molecules were removed to strictly limit the dataset to closed-shell systems.
    \item \textbf{Deduplication:} Duplicate entries were removed within each subset.
    The \texttt{mapped\_isomeric\_smiles} string was utilized as the primary deduplication key.
    For the TorsionScan and TorsionScanRelax subsets, the \texttt{torsion\_indices} were appended to this key to correctly distinguish unique torsional profiles.
    \item \textbf{Geometry Convergence:} Entries with unconverged geometry optimizations were discarded from the HessianRelax and TorsionScanRelax subsets.
    A configuration was flagged and removed if its maximum atomic force norm exceeded 0.2 \si{\eV\per\angstrom} (approximately 4.6 \si{\kcal\per\mol\per\angstrom}).
    For the TorsionScanRelax subset, the fifth-largest atomic force norm was evaluated instead of the absolute maximum because the four constrained atoms naturally exhibit artificial forces due to the applied geometric constraints.
    \item \textbf{Torsional Consistency:} Torsional dihedral angles were rigorously checked for the TorsionScan and TorsionScanRelax subsets.
    First, a trajectory was rejected if the final constrained angle deviated from the initial target by more than 1.0$^\circ$.
    Second, the correctness of the labeled torsion indices was verified.
    Data were discarded if the sampled constraint angles for specific torsion indices varied by less than 2.0$^\circ$.
\end{itemize}

\section{Data}
\subsection{Format}

The dataset consists of five subsets: Hessian, HessianRelax, TorsionScan, TorsionScanRelax, and MBIS.
It is distributed in a hybrid format utilizing both comma-separated values (CSV) and Hierarchical Data Format version 5 (HDF5) files.
Each data entry is uniquely identified by a universally unique identifier (UUID).
For the Hessian, HessianRelax, and MBIS subsets, the UUID is deterministically generated based on the molecule.
For the TorsionScan and TorsionScanRelax subsets, the UUID is deterministically generated based on both the molecule and the torsion indices.
The CSV files provide metadata, including the UUID, mapped non-isomeric and isomeric SMILES strings, and the corresponding HDF5 file paths.
For specific subsets, additional metadata such as the number of relaxation steps or torsion atom indices are also included in the CSV files.

To maintain consistency, all physical quantities are reported in standard units.
Coordinates are given in angstroms (\si{\angstrom}).
Energies, forces, and Hessian matrices are reported in \si{\kcal\per\mol}, \si{\kcal\per\mol\per\angstrom}, and \si{\kcal\per\mol\per\angstrom\squared}, respectively.
Atomic charges are provided in elementary charge units (\si{\e}), while dipoles and quadrupoles are in \si{\e\angstrom} and \si{\e\angstrom\squared}.
Atomic volumes are expressed in \si{\angstrom\cubed}, and the inverse width parameters of the MBIS Slater functions are in \si{\per\angstrom}.

The HDF5 files store the detailed quantum mechanical properties and geometric data, organized hierarchically with the UUID as the root group for each data entry.
Common attributes shared across all subsets include the mapped SMILES strings and atomic numbers.
The Hessian subset contains optimized coordinates and the corresponding $3N \times 3N$ Hessian matrices, where $N$ denotes the number of atoms in the molecule.
For the relaxation trajectories in the HessianRelax and TorsionScanRelax subsets, step-wise geometries, energies, and forces are recorded.
The TorsionScan and TorsionScanRelax subsets explicitly define the four atom indices involved in the constrained dihedral angle.
Finally, the MBIS subset provides atomic coordinates along with derived properties, including atomic volumes, charges, dipoles, quadrupoles, and parameters for the Minimal Basis Iterative Stockholder Slater functions.
A comprehensive specification of the CSV columns and HDF5 internal structures for each subset is provided in the Appendix.

\subsection{Dataset Composition and Structural Statistics}
To demonstrate the broad chemical and structural diversity captured by this work, we detail the compositional demographics and computational statistics of the dataset below.
Table~\ref{tab:dataset_overview} provides a summary of the dataset sizes and computational details for each subset.

\begin{table}[H]
  \centering
  \caption{
    Overview of the THEMol dataset subsets.
    This table summarizes the level of theory, the total number of entries, and supplementary computational metrics for each subset.
    An \textit{Entry} corresponds to a unique universally unique identifier (UUID).
    For the Hessian, HessianRelax, and MBIS subsets, each entry represents a unique molecule.
    For the TorsionScan and TorsionScanRelax subsets, each entry represents a unique combination of a molecule and specific torsion indices.
    Furthermore, \textit{Molecules} and \textit{Constraints} designate the number of unique molecular graphs and the total number of torsion angles, respectively.
    For the HessianRelax and TorsionScanRelax subsets, \textit{Steps} indicates the total number of recorded geometry optimization steps across all relaxation trajectories.
  }
  \label{tab:dataset_overview}
  \resizebox{\textwidth}{!}{
  \begin{tabular}{llrl}
    \toprule
    Subset & Level of Theory & Entries & Supplementary Metrics \\
    \midrule
    Hessian & B3LYP-D3(BJ)/DZVP & 3,102,537 & --- \\
    HessianRelax & B3LYP-D3(BJ)/DZVP & 4,811,722 & 281,123,880 steps \\
    TorsionScan & B3LYP-D3(BJ)/DZVP & 4,192,791 & 2,436,985 molecules; 93,994,576 constraints \\
    TorsionScanRelax & B3LYP-D3(BJ)/DZVP & 4,914,677 & 3,090,560 molecules; 110,235,160 constraints; 2,993,685,868 steps \\
    MBIS & PBE0/def2-TZVPD (DZVP used only for I atoms) & 3,082,151 & --- \\
    \bottomrule
  \end{tabular}
  }
\end{table}

\begin{figure}[H]
  \centering
  \includegraphics[width=\textwidth]{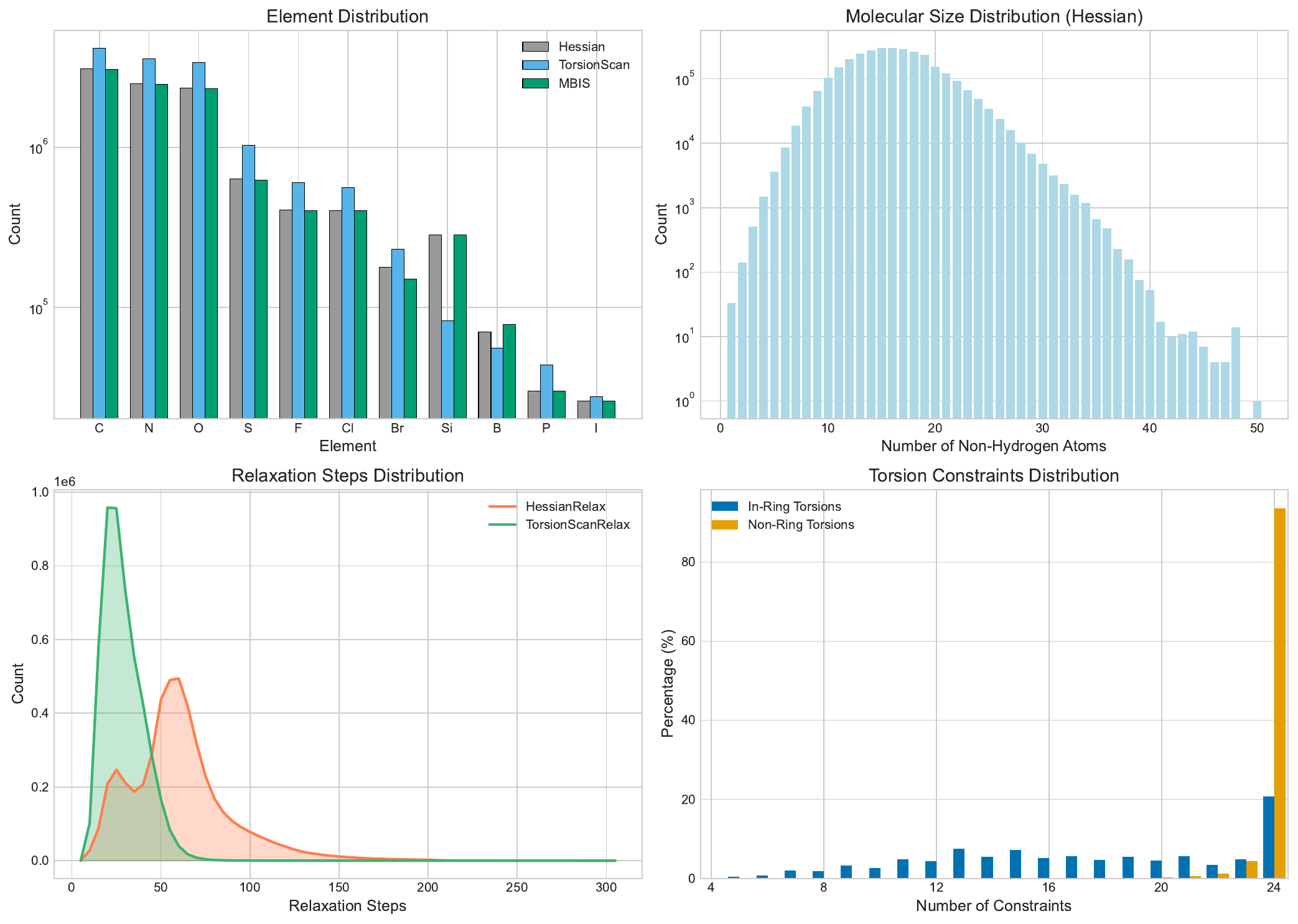}
  \caption{
    Key distributions across the dataset.
    Top left: Element counts at the molecule level for selected subsets (excluding hydrogen).
    Top right: Molecular size distribution (using the Hessian subset as a representative example), measured by the number of non-hydrogen atoms.
    Bottom left: Distribution of relaxation steps for the HessianRelax and TorsionScanRelax subsets.
    Bottom right: Percentage distribution of valid constraints during in-ring and non-ring torsion scans in the TorsionScan subset.
  }
  \label{fig:statistics_histograms}
\end{figure}

The top left panel of Figure~\ref{fig:statistics_histograms} presents the molecular-level element distributions across three key subsets: Hessian, TorsionScan, and MBIS.
As illustrated, carbon naturally emerges as the most prevalent element, appearing in nearly all molecules and accounting for over 70\% of all atoms (shown in Table~\ref{tab:element_atom} of the Appendix).
Other essential organic elements, such as nitrogen and oxygen, similarly exhibit massive representation.
Beyond these foundational elements, the dataset achieves remarkable chemical diversity by thoroughly sampling heavier and less common atomic species.
Sulfur and halogens are highly represented; for instance, sulfur appears in over 600,000 molecules within the Hessian subset alone.
Crucially, even for the heaviest and rarest element within the dataset's scope, iodine (I), over 25,000 unique molecular structures are provided across these subsets.
This unprecedented volume of data for both foundational and heavier main-group elements ensures that models trained on this dataset can robustly generalize to complex organic chemical spaces.
Finally, further detailed element distributions at both the molecular and atomic levels are provided in the Appendix.

Figure~\ref{fig:statistics_histograms} also illustrates three other fundamental structural and computational distributions within the dataset.
The top right panel displays the molecular size distribution, measured by the total number of heavy (non-hydrogen) atoms.
The Hessian subset is shown as a representative example, as the size distribution remains highly consistent across all other subsets.
It shows that the majority of molecules in the dataset have between 8 and 25 heavy atoms.
The bottom left panel contrasts the distribution of required relaxation steps for the HessianRelax and TorsionScanRelax subsets.
The trajectories in the TorsionScanRelax subset converge in significantly fewer steps.
This behavior is expected because the initial configurations for torsional scans are generated by rotating a specific dihedral angle within an already optimized geometry.
Finally, the bottom right panel contrasts the final number of successfully converged geometries (valid constraints) for in-ring versus non-ring torsion scans.
In-ring scans naturally yield fewer valid constraints in the data because the inherent structural rigidity of the cyclic system physically prevents a complete 360-degree rotation.

\section{Discussion and Conclusion}

In this work, we introduce THEMol, a comprehensive open-source collection of quantum mechanical properties tailored for drug-like molecules.

We highlight several known limitations to guide responsible use.
First, the dataset was accumulated over a span of four years using two electronic structure engines, namely Q-Chem on CPU in the early phase and GPU4PySCF after it matured.
Minor methodological differences between engines and default settings may introduce small inconsistencies in edge cases.
Second, the current sampling contains very few species in hypervalent coordination environments with more than four bonded neighbors.
As a result, certain motifs such as octahedral phosphorus (for example, nominal sp3d2) are underrepresented.
We plan to expand coverage to include these coordination states in future updates.
Third, we rely on RDKit to generate isomeric SMILES from 3D coordinates for deduplication and post-processing.
In rare cases with challenging stereochemistry or near-degenerate conformations, automated assignment can be imperfect.
When exact stereochemical labels are critical, we recommend that users recompute CIP assignments directly from the provided coordinates as a safety check.
Fourth, we adopt the general D3(BJ) dispersion parameters with the B3LYP functional.
Some users prefer alternative parameter sets tailored specifically to the DZVP basis, which can lead to minor numerical differences.
We do not expect material impact on most applications, but reproducibility at the sub-kcal level may require adopting the same settings.
Fifth, for iodine-containing molecules, the sharply peaked electron density near the nucleus can challenge MBIS partitioning.
This may occasionally reduce accuracy or lead to convergence difficulties for density-derived properties.
We advise applying routine sanity checks when using the MBIS subset for heavy-halogen systems.

Looking forward, we see several directions to further enhance utility and coverage.
Taken together, THEMol provides large-scale, diverse, and systematically organized quantum mechanical data that enable training and rigorous evaluation of both molecular mechanics and machine learning potentials.
We anticipate that THEMol will serve as a highly valuable foundational resource for computational chemistry and the broader AI for Science community.
Ultimately, we hope this dataset will empower the development of accurate, robust, and transferable models across a wide range of organic molecules and applications.

\section{Data Availability}

The THEMol dataset, including all subsets and associated metadata, is publicly accessible via the Hugging Face repository at \url{https://huggingface.co/datasets/ByteDance-Seed/THEMol}.
Additionally, a dedicated GitHub repository containing validation utilities, example data loaders, and statistical scripts is provided at \url{https://github.com/ByteDance-Seed/THEMol} to facilitate data processing and analysis.

\section{Acknowledgement}
ByteDance Inc. holds intellectual property rights pertinent to the research presented herein.

\clearpage

\bibliographystyle{unsrtnat}
\bibliography{main}

\begin{thebibliography}{50}
\providecommand{\natexlab}[1]{#1}
\providecommand{\url}[1]{\texttt{#1}}
\expandafter\ifx\csname urlstyle\endcsname\relax
  \providecommand{\doi}[1]{doi: #1}\else
  \providecommand{\doi}{doi: \begingroup \urlstyle{rm}\Url}\fi

\bibitem[Ramakrishnan et~al.(2014)Ramakrishnan, Dral, Rupp, and {von Lilienfeld}]{ramakrishnanQuantumChemistryStructures2014}
Raghunathan Ramakrishnan, Pavlo~O. Dral, Matthias Rupp, and O.~Anatole {von Lilienfeld}.
\newblock Quantum chemistry structures and properties of 134 kilo molecules.
\newblock \emph{Scientific Data}, 1\penalty0 (1):\penalty0 140022, 2014.
\newblock ISSN 2052-4463.
\newblock \doi{10.1038/sdata.2014.22}.

\bibitem[Nakata et~al.(2020)Nakata, Shimazaki, Hashimoto, and Maeda]{nakataPubChemQCPM6Data2020}
Maho Nakata, Tomomi Shimazaki, Masatomo Hashimoto, and Toshiyuki Maeda.
\newblock {{PubChemQC PM6}}: {{Data Sets}} of 221 {{Million Molecules}} with {{Optimized Molecular Geometries}} and {{Electronic Properties}}.
\newblock \emph{Journal of Chemical Information and Modeling}, 60\penalty0 (12):\penalty0 5891--5899, 2020.
\newblock ISSN 1549-9596.
\newblock \doi{10.1021/acs.jcim.0c00740}.

\bibitem[Nakata and Maeda(2023)]{nakataPubChemQCB3LYP631G2023}
Maho Nakata and Toshiyuki Maeda.
\newblock {{PubChemQC B3LYP}}/6-{{31G}}*//{{PM6 Data Set}}: {{The Electronic Structures}} of 86 {{Million Molecules Using B3LYP}}/6-{{31G}}* {{Calculations}}.
\newblock \emph{Journal of Chemical Information and Modeling}, 63\penalty0 (18):\penalty0 5734--5754, 2023.
\newblock ISSN 1549-9596.
\newblock \doi{10.1021/acs.jcim.3c00899}.

\bibitem[Ganscha et~al.(2025)Ganscha, Unke, Ahlin, Maennel, Kashubin, and M{\"u}ller]{ganscha2025qcml}
Stefan Ganscha, Oliver~T Unke, Daniel Ahlin, Hartmut Maennel, Sergii Kashubin, and Klaus-Robert M{\"u}ller.
\newblock The qcml dataset, quantum chemistry reference data from 33.5m dft and 14.7b semi-empirical calculations.
\newblock \emph{Scientific Data}, 12\penalty0 (1):\penalty0 406, 2025.
\newblock ISSN 2052-4463.
\newblock \doi{10.1038/s41597-025-04720-7}.

\bibitem[Smith et~al.(2017{\natexlab{a}})Smith, Isayev, and Roitberg]{smithANI1DataSet2017}
Justin~S. Smith, Olexandr Isayev, and Adrian~E. Roitberg.
\newblock {{ANI-1}}, {{A}} data set of 20 million calculated off-equilibrium conformations for organic molecules.
\newblock \emph{Scientific Data}, 4\penalty0 (1):\penalty0 170193, 2017{\natexlab{a}}.
\newblock ISSN 2052-4463.
\newblock \doi{10.1038/sdata.2017.193}.

\bibitem[Smith et~al.(2017{\natexlab{b}})Smith, Isayev, and Roitberg]{smith2017ani1}
Justin~S Smith, Olexandr Isayev, and Adrian~E Roitberg.
\newblock Ani-1: an extensible neural network potential with dft accuracy at force field computational cost.
\newblock \emph{Chemical Science}, 8\penalty0 (4):\penalty0 3192--3203, 2017{\natexlab{b}}.

\bibitem[Smith et~al.(2020)Smith, Zubatyuk, Nebgen, Lubbers, Barros, Roitberg, Isayev, and Tretiak]{smith2020ani1x}
Justin~S Smith, Roman Zubatyuk, Benjamin Nebgen, Nicholas Lubbers, Kipton Barros, Adrian~E Roitberg, Olexandr Isayev, and Sergei Tretiak.
\newblock The ani-1ccx and ani-1x data sets, coupled-cluster and density functional theory properties for molecules.
\newblock \emph{Scientific Data}, 7\penalty0 (1):\penalty0 134, 2020.
\newblock ISSN 2052-4463.
\newblock \doi{10.1038/s41597-020-0473-z}.

\bibitem[Verstraelen et~al.(2016)Verstraelen, Vandenbrande, Heidar-Zadeh, Vanduyfhuys, Van~Speybroeck, Waroquier, and Ayers]{verstraelen2016minimal}
Toon Verstraelen, Steven Vandenbrande, Farnaz Heidar-Zadeh, Louis Vanduyfhuys, Veronique Van~Speybroeck, Michel Waroquier, and Paul~W Ayers.
\newblock Minimal basis iterative stockholder: atoms in molecules for force-field development.
\newblock \emph{Journal of Chemical Theory and Computation}, 12\penalty0 (8):\penalty0 3894--3912, 2016.

\bibitem[Pulido et~al.(2024)Pulido, Macaya, and Vohringer-Martinez]{pulido2024nonbonded}
Jorge Pulido, Luis Macaya, and Esteban Vohringer-Martinez.
\newblock Nonbonded force field parameters from mbis partitioning of the molecular electron density improve thermophysical properties prediction of organic liquids.
\newblock \emph{Journal of Chemical \& Engineering Data}, 69\penalty0 (9):\penalty0 2917--2926, 2024.

\bibitem[Smith et~al.(2019)Smith, Nebgen, Zubatyuk, Lubbers, Devereux, Barros, Tretiak, Isayev, and Roitberg]{smithApproachingCoupledCluster2019}
Justin~S. Smith, Benjamin~T. Nebgen, Roman Zubatyuk, Nicholas Lubbers, Christian Devereux, Kipton Barros, Sergei Tretiak, Olexandr Isayev, and Adrian~E. Roitberg.
\newblock Approaching coupled cluster accuracy with a general-purpose neural network potential through transfer learning.
\newblock \emph{Nature Communications}, 10\penalty0 (1):\penalty0 2903, 2019.
\newblock ISSN 2041-1723.
\newblock \doi{10.1038/s41467-019-10827-4}.

\bibitem[Devereux et~al.(2020)Devereux, Smith, Huddleston, Barros, Zubatyuk, Isayev, and Roitberg]{devereux2020ani2x}
Christian Devereux, Justin~S Smith, Kipton~K Huddleston, Kipton Barros, Roman Zubatyuk, Olexandr Isayev, and Adrian~E Roitberg.
\newblock Extending the applicability of the ani deep learning molecular potential to sulfur and halogens.
\newblock \emph{Journal of Chemical Theory and Computation}, 16\penalty0 (7):\penalty0 4192--4202, 2020.

\bibitem[Zhang et~al.(2025)Zhang, Zubatyuk, Yang, Roitberg, and Isayev]{zhangANI1xBBANIBasedReactive2025}
Shuhao Zhang, Roman Zubatyuk, Yinuo Yang, Adrian Roitberg, and Olexandr Isayev.
\newblock {{ANI-1xBB}}: {{An ANI-Based Reactive Potential}} for {{Small Organic Molecules}}.
\newblock \emph{Journal of Chemical Theory and Computation}, 21\penalty0 (9):\penalty0 4365--4374, 2025.
\newblock ISSN 1549-9618.
\newblock \doi{10.1021/acs.jctc.5c00347}.

\bibitem[Zubatyuk et~al.(2019)Zubatyuk, Smith, Leszczynski, and Isayev]{zubatyuk2019aimnet}
Roman Zubatyuk, Justin~S Smith, Jerzy Leszczynski, and Olexandr Isayev.
\newblock Accurate and transferable multitask prediction of chemical properties with an atoms-in-molecules neural network.
\newblock \emph{Science Advances}, 5\penalty0 (8):\penalty0 eaav6490, 2019.
\newblock \doi{10.1126/sciadv.aav6490}.

\bibitem[Anstine et~al.(2025{\natexlab{a}})Anstine, Zubatyuk, and Isayev]{anstineAIMNet2NeuralNetwork2025}
Dylan~M. Anstine, Roman Zubatyuk, and Olexandr Isayev.
\newblock {{AIMNet2}}: A neural network potential to meet your neutral, charged, organic, and elemental-organic needs.
\newblock \emph{Chemical Science}, 16\penalty0 (23):\penalty0 10228--10244, 2025{\natexlab{a}}.
\newblock ISSN 2041-6539.
\newblock \doi{10.1039/D4SC08572H}.

\bibitem[Kalita et~al.(2026)Kalita, Zubatyuk, Anstine, Bergeler, Settels, Stork, Spicher, and Isayev]{kalitaAIMNet2NSETransferableReactive2026a}
Bhupalee Kalita, Roman Zubatyuk, Dylan~M. Anstine, Maike Bergeler, Volker Settels, Conrad Stork, Sebastian Spicher, and Olexandr Isayev.
\newblock {{AIMNet2-NSE}}: {{A Transferable Reactive Neural Network Potential}} for {{Open-Shell Chemistry}}.
\newblock \emph{Angewandte Chemie International Edition}, 65\penalty0 (5):\penalty0 e16763, 2026.
\newblock ISSN 1521-3773.
\newblock \doi{10.1002/anie.202516763}.

\bibitem[Anstine et~al.(2025{\natexlab{b}})Anstine, Zhao, Zubatiuk, Isayev, et~al.]{anstine2025aimnet2rxn}
Dylan~M Anstine, Qiyuan Zhao, Roman Zubatiuk, Olexandr Isayev, et~al.
\newblock Aimnet2-rxn: A machine learned potential for generalized reaction modeling on a millions-of-pathways scale.
\newblock \emph{ChemRxiv}, 2025{\natexlab{b}}.
\newblock \doi{10.26434/chemrxiv-2025-hpdmg}.

\bibitem[Levine et~al.(2025{\natexlab{a}})Levine, Shuaibi, Spotte-Smith, Taylor, Hasyim, Michel, Batatia, Cs{\'a}nyi, Dzamba, Eastman, et~al.]{levine2025omol25}
Daniel~S Levine, Muhammed Shuaibi, Evan Walter~Clark Spotte-Smith, Michael~G Taylor, Muhammad~R Hasyim, Kyle Michel, Ilyes Batatia, G{\'a}bor Cs{\'a}nyi, Misko Dzamba, Peter Eastman, et~al.
\newblock The open molecules 2025 (omol25) dataset, evaluations, and models.
\newblock \emph{arXiv preprint arXiv:2505.08762}, 2025{\natexlab{a}}.

\bibitem[Levine et~al.(2025{\natexlab{b}})Levine, Liesen, Chua, et~al.]{levine2025opoly26}
Daniel~S Levine, Nicholas~T Liesen, Lauren Chua, et~al.
\newblock The open polymers 2026 (opoly26) dataset and evaluations.
\newblock \emph{arXiv preprint arXiv:2512.23117}, 2025{\natexlab{b}}.

\bibitem[Kabylda et~al.(2026)Kabylda, {Su{\'a}rez-Dou}, Davoine, Br{\"u}nig, and Tkatchenko]{kabyldaQCellComprehensiveQuantumMechanical2026b}
Adil Kabylda, Sergio {Su{\'a}rez-Dou}, Nils Davoine, Florian~N. Br{\"u}nig, and Alexandre Tkatchenko.
\newblock {{QCell}}: {{Comprehensive Quantum-Mechanical Dataset Spanning Diverse Biomolecular Fragments}}.
\newblock \emph{AI for Science}, 2026.
\newblock ISSN 3050-287X.
\newblock \doi{10.1088/3050-287X/ae5267}.

\bibitem[Eastman et~al.(2023)Eastman, Behara, Dotson, Galvelis, Herr, Horton, Mao, Chodera, Pritchard, Wang, De~Fabritiis, and Markland]{eastman2023spice}
Peter Eastman, Pavan~Kumar Behara, David~L Dotson, Raimondas Galvelis, John~E Herr, Josh~T Horton, Yuezhi Mao, John~D Chodera, Benjamin~P Pritchard, Yuanqing Wang, Gianni De~Fabritiis, and Thomas~E Markland.
\newblock Spice, a dataset of drug-like molecules and peptides for training machine learning potentials.
\newblock \emph{Scientific Data}, 10\penalty0 (1):\penalty0 11, 2023.
\newblock ISSN 2052-4463.
\newblock \doi{10.1038/s41597-022-01882-6}.

\bibitem[Axelrod and G{\'o}mez-Bombarelli(2022)]{axelrod2022geom}
Simon Axelrod and Rafael G{\'o}mez-Bombarelli.
\newblock Geom, energy-annotated molecular conformations for property prediction and molecular generation.
\newblock \emph{Scientific Data}, 9\penalty0 (1):\penalty0 185, 2022.
\newblock ISSN 2052-4463.
\newblock \doi{10.1038/s41597-022-01288-4}.

\bibitem[Isert et~al.(2022)Isert, Atz, Jim{\'e}nez-Luna, and Schneider]{isert2022qmugs}
Clemens Isert, Kenneth Atz, Jos{\'e} Jim{\'e}nez-Luna, and Gisbert Schneider.
\newblock Qmugs, quantum mechanical properties of drug-like molecules.
\newblock \emph{Scientific Data}, 9\penalty0 (1):\penalty0 273, 2022.

\bibitem[Schreiner et~al.(2022)Schreiner, Bhowmik, Vegge, Busk, and Winther]{schreinerTransition1xDatasetBuilding2022}
Mathias Schreiner, Arghya Bhowmik, Tejs Vegge, Jonas Busk, and Ole Winther.
\newblock Transition1x - a dataset for building generalizable reactive machine learning potentials.
\newblock \emph{Scientific Data}, 9\penalty0 (1):\penalty0 779, 2022.
\newblock ISSN 2052-4463.
\newblock \doi{10.1038/s41597-022-01870-w}.

\bibitem[Zhao et~al.(2025)Zhao, Han, Cui, et~al.]{zhao2025horm}
Qiyuan Zhao, Yunhong Han, Taoyong Cui, et~al.
\newblock Horm: A large scale molecular hessian database for optimizing reactive machine learning interatomic potentials.
\newblock \emph{arXiv preprint arXiv:2505.12447}, 2025.

\bibitem[Donchev et~al.(2021)Donchev, Taube, Decolvenaere, Hargus, McGibbon, Law, Gregersen, Li, Palmo, Siva, Bergdorf, Klepeis, and Shaw]{donchev2021des}
Alexander~G Donchev, Andrew~G Taube, Elizabeth Decolvenaere, Cory Hargus, Robert~T McGibbon, Ka-Hei Law, Brent~A Gregersen, Je-Luen Li, Kim Palmo, Karthik Siva, Michael Bergdorf, John~L Klepeis, and David~E Shaw.
\newblock Quantum chemical benchmark databases of gold-standard dimer interaction energies.
\newblock \emph{Scientific Data}, 8\penalty0 (1):\penalty0 55, 2021.

\bibitem[Burns et~al.(2017)Burns, Faver, Zheng, Marshall, Smith, Vanommeslaeghe, MacKerell, Merz, and Sherrill]{burnsBioFragmentDatabaseBFDb2017}
Lori~A. Burns, John~C. Faver, Zheng Zheng, Michael~S. Marshall, Daniel G.~A. Smith, Kenno Vanommeslaeghe, Alexander~D. MacKerell, Kenneth~M. Merz, and C.~David Sherrill.
\newblock The {{BioFragment Database}} ({{BFDb}}): {{An}} open-data platform for computational chemistry analysis of noncovalent interactions.
\newblock \emph{The Journal of Chemical Physics}, 147\penalty0 (16):\penalty0 161727, 2017.
\newblock ISSN 0021-9606.
\newblock \doi{10.1063/1.5001028}.

\bibitem[{\v R}ez{\'a}{\v c}(2020)]{rezacNonCovalentInteractionsAtlas2020a}
Jan {\v R}ez{\'a}{\v c}.
\newblock Non-{{Covalent Interactions Atlas Benchmark Data Sets}}: {{Hydrogen Bonding}}.
\newblock \emph{Journal of Chemical Theory and Computation}, 16\penalty0 (4):\penalty0 2355--2368, 2020.
\newblock ISSN 1549-9618.
\newblock \doi{10.1021/acs.jctc.9b01265}.

\bibitem[Wang et~al.(2023)Wang, He, Li, Shao, and Liu]{wangAIMDChigExploringConformational2023}
Tong Wang, Xinheng He, Mingyu Li, Bin Shao, and Tie-Yan Liu.
\newblock {{AIMD-Chig}}: {{Exploring}} the conformational space of a 166-atom protein {{Chignolin}} with ab initio molecular dynamics.
\newblock \emph{Scientific Data}, 10\penalty0 (1):\penalty0 549, 2023.
\newblock ISSN 2052-4463.
\newblock \doi{10.1038/s41597-023-02465-9}.

\bibitem[Williams et~al.(2025)Williams, Kabalan, Stojanovic, Z{\'o}lyomi, and {Pyzer-Knapp}]{williamsHessianQM9Quantum2025}
Nicholas~J. Williams, Lara Kabalan, Ljiljana Stojanovic, Viktor Z{\'o}lyomi, and Edward~O. {Pyzer-Knapp}.
\newblock Hessian {{QM9}}: {{A}} quantum chemistry database of molecular {{Hessians}} in implicit solvents.
\newblock \emph{Scientific Data}, 12\penalty0 (1):\penalty0 9, 2025.
\newblock ISSN 2052-4463.
\newblock \doi{10.1038/s41597-024-04361-2}.

\bibitem[Zapata~Trujillo and McKemmish(2022)]{zapatatrujilloVIBFREQ1295NewDatabase2022}
Juan~C. Zapata~Trujillo and Laura~K. McKemmish.
\newblock {{VIBFREQ1295}}: {{A New Database}} for {{Vibrational Frequency Calculations}}.
\newblock \emph{The Journal of Physical Chemistry A}, 126\penalty0 (25):\penalty0 4100--4122, 2022.
\newblock ISSN 1089-5639.
\newblock \doi{10.1021/acs.jpca.2c01438}.

\bibitem[Behara et~al.(2024)Behara, Jang, Horton, Gokey, Dotson, Boothroyd, Bayly, Cole, Wang, and Mobley]{beharaBenchmarkDFT2024}
Pavan~Kumar Behara, Hyesu Jang, Joshua~T. Horton, Trevor Gokey, David~L. Dotson, Simon Boothroyd, Christopher~I. Bayly, Daniel~J. Cole, Lee-Ping Wang, and David~L. Mobley.
\newblock Benchmarking quantum mechanical levels of theory for valence parametrization in force fields.
\newblock \emph{The Journal of Physical Chemistry B}, 128\penalty0 (32):\penalty0 7888--7902, 2024.
\newblock \doi{10.1021/acs.jpcb.4c03167}.
\newblock PMID: 39087913.

\bibitem[Zheng et~al.(2025)Zheng, Wang, Han, Xia, Xu, Zhan, Liu, Chen, Wang, Wu, Gong, and Yan]{zhengByteFF24}
Tianze Zheng, Ailun Wang, Xu~Han, Yu~Xia, Xingyuan Xu, Jiawei Zhan, Yu~Liu, Yang Chen, Zhi Wang, Xiaojie Wu, Sheng Gong, and Wen Yan.
\newblock Data-driven parametrization of molecular mechanics force fields for expansive chemical space coverage.
\newblock \emph{Chem. Sci.}, 16:\penalty0 2730--2740, 2025.
\newblock \doi{10.1039/D4SC06640E}.

\bibitem[Chambers et~al.(2013)Chambers, Davies, Gaulton, Hersey, Velankar, Petryszak, Hastings, Bellis, McGlinchey, and Overington]{chambers2013unichem}
Jon Chambers, Mark Davies, Anna Gaulton, Anne Hersey, Sameer Velankar, Robert Petryszak, Janna Hastings, Louisa Bellis, Shaun McGlinchey, and John~P Overington.
\newblock Unichem: a unified chemical structure cross-referencing and identifier tracking system.
\newblock \emph{Journal of cheminformatics}, 5\penalty0 (1):\penalty0 3, 2013.

\bibitem[Lu et~al.(2021)Lu, Wu, Ghoreishi, Chen, Wang, Damm, Ross, Dahlgren, Russell, Von~Bargen, Abel, Friesner, and Harder]{LuSFE2021}
Chao Lu, Chuanjie Wu, Delaram Ghoreishi, Wei Chen, Lingle Wang, Wolfgang Damm, Gregory~A. Ross, Markus~K. Dahlgren, Ellery Russell, Christopher~D. Von~Bargen, Robert Abel, Richard~A. Friesner, and Edward~D. Harder.
\newblock {OPLS4}: Improving force field accuracy on challenging regimes of chemical space.
\newblock \emph{Journal of Chemical Theory and Computation}, 17\penalty0 (7):\penalty0 4291--4300, 2021.
\newblock ISSN 1549-9618.
\newblock \doi{10.1021/acs.jctc.1c00302}.

\bibitem[Riccardi et~al.(2021)Riccardi, Bazyleva, Paulechka, Diky, Magee, Kazakov, Townsend, and Muzny]{ThermoMLArchivevdw2021}
Demian Riccardi, Ala Bazyleva, Eugene Paulechka, Vladimir Diky, Josepha~W. Magee, Andrei~F. Kazakov, Scott~A. Townsend, and Chris~D. Muzny.
\newblock {ThermoML} data archive, 2021.
\newblock URL \url{https://trc.nist.gov/ThermoML/}.
\newblock Accessed: 2025-03-30.

\bibitem[Wang et~al.(2015)Wang, Wu, Deng, Kim, Pierce, Krilov, Lupyan, Robinson, Dahlgren, Greenwood, Romero, Masse, Knight, Steinbrecher, Beuming, Damm, Harder, Sherman, Brewer, Wester, Murcko, Frye, Farid, Lin, Mobley, Jorgensen, Berne, Friesner, and Abel]{Wang2015}
Lingle Wang, Yujie Wu, Yuqing Deng, Byungchan Kim, Levi Pierce, Goran Krilov, Dmitry Lupyan, Shaughnessy Robinson, Markus~K. Dahlgren, Jeremy Greenwood, Donna~L. Romero, Craig Masse, Jennifer~L. Knight, Thomas Steinbrecher, Thijs Beuming, Wolfgang Damm, Ed~Harder, Woody Sherman, Mark Brewer, Ron Wester, Mark Murcko, Leah Frye, Ramy Farid, Teng Lin, David~L. Mobley, William~L. Jorgensen, Bruce~J. Berne, Richard~A. Friesner, and Robert Abel.
\newblock Accurate and reliable prediction of relative ligand binding potency in prospective drug discovery by way of a modern free-energy calculation protocol and force field.
\newblock \emph{Journal of the American Chemical Society}, 137\penalty0 (7):\penalty0 2695--2703, 2015.
\newblock ISSN 0002-7863.
\newblock \doi{10.1021/ja512751q}.

\bibitem[Wang et~al.(2017)Wang, Deng, Wu, Kim, LeBard, Wandschneider, Beachy, Friesner, and Abel]{Wang2017}
Lingle Wang, Yuqing Deng, Yujie Wu, Byungchan Kim, David~N. LeBard, Dan Wandschneider, Mike Beachy, Richard~A. Friesner, and Robert Abel.
\newblock Accurate modeling of scaffold hopping transformations in drug discovery.
\newblock \emph{Journal of Chemical Theory and Computation}, 13\penalty0 (1):\penalty0 42--54, 2017.
\newblock ISSN 1549-9618.
\newblock \doi{10.1021/acs.jctc.6b00991}.

\bibitem[Schindler et~al.(2020)Schindler, Baumann, Blum, B{\"o}se, Buchstaller, Burgdorf, Cappel, Chekler, Czodrowski, Dorsch, Eguida, Follows, Fuch{\ss}, Gr{\"a}dler, Gunera, Johnson, Jorand~Lebrun, Karra, Klein, Knehans, Koetzner, Krier, Leiendecker, Leuthner, Li, Mochalkin, Musil, Neagu, Rippmann, Schiemann, Schulz, Steinbrecher, Tanzer, {Unzue Lopez}, {Viacava Follis}, Wegener, and Kuhn]{Schindler2020}
Christina E.~M. Schindler, Hannah Baumann, Andreas Blum, Dietrich B{\"o}se, Hans-Peter Buchstaller, Lars Burgdorf, Daniel Cappel, Eugene Chekler, Paul Czodrowski, Dieter Dorsch, Merveille K.~I. Eguida, Bruce Follows, Thomas Fuch{\ss}, Ulrich Gr{\"a}dler, Jakub Gunera, Theresa Johnson, Catherine Jorand~Lebrun, Srinivasa Karra, Markus Klein, Tim Knehans, Lisa Koetzner, Mireille Krier, Matthias Leiendecker, Birgitta Leuthner, Liwei Li, Igor Mochalkin, Djordje Musil, Constantin Neagu, Friedrich Rippmann, Kai Schiemann, Robert Schulz, Thomas Steinbrecher, Eva-Maria Tanzer, Andrea {Unzue Lopez}, Ariele {Viacava Follis}, Ansgar Wegener, and Daniel Kuhn.
\newblock Large-scale assessment of binding free energy calculations in active drug discovery projects.
\newblock \emph{Journal of Chemical Information and Modeling}, 60\penalty0 (11):\penalty0 5457--5474, 2020.
\newblock ISSN 1549-9596.
\newblock \doi{10.1021/acs.jcim.0c00900}.

\bibitem[Jinsong et~al.(2024)Jinsong, Qifeng, Xing, Hao, and Wang]{jinsong2024molecular}
Shao Jinsong, Jia Qifeng, Chen Xing, Yajie Hao, and Li~Wang.
\newblock Molecular fragmentation as a crucial step in the ai-based drug development pathway.
\newblock \emph{Communications Chemistry}, 7\penalty0 (1):\penalty0 20, 2024.

\bibitem[Roos et~al.(2019)Roos, Wu, Damm, Reboul, Stevenson, Lu, Dahlgren, Mondal, Chen, Wang, Abel, Friesner, and Harder]{roosOPLS3eExtendingForce2019}
Katarina Roos, Chuanjie Wu, Wolfgang Damm, Mark Reboul, James~M. Stevenson, Chao Lu, Markus~K. Dahlgren, Sayan Mondal, Wei Chen, Lingle Wang, Robert Abel, Richard~A. Friesner, and Edward~D. Harder.
\newblock {{OPLS3e}}: {{Extending Force Field Coverage}} for {{Drug-Like Small Molecules}}.
\newblock \emph{Journal of Chemical Theory and Computation}, 15\penalty0 (3):\penalty0 1863--1874, 2019.
\newblock ISSN 1549-9618.
\newblock \doi{10.1021/acs.jctc.8b01026}.

\bibitem[Shelley et~al.(2007)Shelley, Cholleti, Frye, Greenwood, Timlin, and Uchimaya]{shelley2007epik}
John~C Shelley, Anuradha Cholleti, Leah~L Frye, Jeremy~R Greenwood, Mathew~R Timlin, and Makoto Uchimaya.
\newblock Epik: a software program for pk a prediction and protonation state generation for drug-like molecules.
\newblock \emph{Journal of computer-aided molecular design}, 21:\penalty0 681--691, 2007.

\bibitem[Wang and Song(2016)]{Wang2016}
Lee-Ping Wang and Chenchen Song.
\newblock Geometry optimization made simple with translation and rotation coordinates.
\newblock \emph{The Journal of Chemical Physics}, 144\penalty0 (21):\penalty0 214108, 2016.
\newblock ISSN 1089-7690.
\newblock \doi{10.1063/1.4952956}.

\bibitem[Stephens et~al.(1994)Stephens, Devlin, Chabalowski, and Frisch]{stephens1994ab}
Philip~J Stephens, Frank~J Devlin, Cary~F Chabalowski, and Michael~J Frisch.
\newblock Ab initio calculation of vibrational absorption and circular dichroism spectra using density functional force fields.
\newblock \emph{The Journal of physical chemistry}, 98\penalty0 (45):\penalty0 11623--11627, 1994.

\bibitem[Dunning~Jr(1989)]{dunning1989gaussian}
Thom~H Dunning~Jr.
\newblock Gaussian basis sets for use in correlated molecular calculations. i. the atoms boron through neon and hydrogen.
\newblock \emph{The Journal of chemical physics}, 90\penalty0 (2):\penalty0 1007--1023, 1989.

\bibitem[Boothroyd et~al.(2023)Boothroyd, Behara, Madin, et~al.]{boothroyd2023openff}
Simon Boothroyd, Pavan~Kumar Behara, Owen Madin, et~al.
\newblock Development and benchmarking of open force field 2.0.0—the sage small molecule force field.
\newblock \emph{ChemRxiv}, 2023.
\newblock \doi{10.26434/chemrxiv-2022-n2z1c-v2}.

\bibitem[Adamo and Barone(1999)]{adamo1999toward}
Carlo Adamo and Vincenzo Barone.
\newblock Toward reliable density functional methods without adjustable parameters: The pbe0 model.
\newblock \emph{The Journal of chemical physics}, 110\penalty0 (13):\penalty0 6158--6170, 1999.

\bibitem[Rappoport and Furche(2010)]{rappoport2010property}
Dmitrij Rappoport and Filipp Furche.
\newblock Property-optimized gaussian basis sets for molecular response calculations.
\newblock \emph{The Journal of chemical physics}, 133\penalty0 (13), 2010.

\bibitem[Medvedev et~al.(2017)Medvedev, Bushmarinov, Sun, Perdew, and Lyssenko]{medvedev2017density}
Michael~G Medvedev, Ivan~S Bushmarinov, Jianwei Sun, John~P Perdew, and Konstantin~A Lyssenko.
\newblock Density functional theory is straying from the path toward the exact functional.
\newblock \emph{Science}, 355\penalty0 (6320):\penalty0 49--52, 2017.

\bibitem[Liang and {Head-Gordon}(2025)]{liangGoldStandardChemicalDatabase2025a}
Jiashu Liang and Martin {Head-Gordon}.
\newblock Gold-{{Standard Chemical Database}} 137 ({{GSCDB137}}): {{A Diverse Set}} of {{Accurate Energy Differences}} for {{Assessing}} and {{Developing Density Functionals}}.
\newblock \emph{Journal of Chemical Theory and Computation}, 2025.
\newblock ISSN 1549-9618.
\newblock \doi{10.1021/acs.jctc.5c01380}.

\bibitem[Hait and Head-Gordon(2018)]{hait2018accurate}
Diptarka Hait and Martin Head-Gordon.
\newblock How accurate is density functional theory at predicting dipole moments? an assessment using a new database of 200 benchmark values.
\newblock \emph{Journal of chemical theory and computation}, 14\penalty0 (4):\penalty0 1969--1981, 2018.

\end{thebibliography}

\clearpage

\beginappendix

\section{Appendix}

\subsection{Detailed Data Format}

This appendix provides a detailed specification of the CSV columns and HDF5 internal file structures for each subset.
Throughout the internal structures, $N$ refers to the number of atoms in the molecule.

\subsubsection{Hessian Subset}

The CSV metadata for the Hessian subset includes the following columns:
\begin{itemize}
  \item \texttt{uuid}
  \item \texttt{mapped\_nonisomeric\_smiles}
  \item \texttt{mapped\_isomeric\_smiles}
  \item \texttt{h5\_file}
\end{itemize}
The corresponding HDF5 file stores data under the root group named by the molecule's UUID.
The internal structure is defined as follows:
\begin{lstlisting}[basicstyle=\ttfamily\small, frame=lines]
/<uuid>/
  mapped_nonisomeric_smiles  (utf-8 string object)
  mapped_isomeric_smiles     (utf-8 string object)
  atomic_numbers             (N, 1) int32
  coords                     (N, 3) float64
  hessian                    (3N, 3N) float64
\end{lstlisting}

\subsubsection{HessianRelax Subset}

The CSV metadata for the HessianRelax subset includes the following columns:
\begin{itemize}
  \item \texttt{uuid}
  \item \texttt{mapped\_nonisomeric\_smiles}
  \item \texttt{mapped\_isomeric\_smiles}
  \item \texttt{num\_steps}
  \item \texttt{h5\_file}
\end{itemize}
The HDF5 internal structure is defined as follows:
\begin{lstlisting}[basicstyle=\ttfamily\small, frame=lines]
/<uuid>/
  mapped_nonisomeric_smiles  (utf-8 string object)
  mapped_isomeric_smiles     (utf-8 string object)
  atomic_numbers             (N, 1) int32
  step 0/
    energy                   scalar float64
    coords                   (N, 3) float64
    forces                   (N, 3) float64
  ...
  step k/
    energy                   scalar float64
    coords                   (N, 3) float64
    forces                   (N, 3) float64
\end{lstlisting}

\subsubsection{TorsionScan Subset}

The CSV metadata for the TorsionScan subset includes the following columns:
\begin{itemize}
  \item \texttt{uuid}
  \item \texttt{mapped\_nonisomeric\_smiles}
  \item \texttt{mapped\_isomeric\_smiles}
  \item \texttt{torsion\_indices}
  \item \texttt{h5\_file}
  \item \texttt{num\_constraints}
\end{itemize}
The HDF5 internal structure is defined as follows:
\begin{lstlisting}[basicstyle=\ttfamily\small, frame=lines]
/<uuid>/
  mapped_nonisomeric_smiles  (utf-8 string object)
  mapped_isomeric_smiles     (utf-8 string object)
  atomic_numbers             (N, 1) int32
  torsion_atom_indices       (4,) int32        # 0-based [i,j,k,l]
  constraint 0/
    energy                   scalar float64
    coords                   (N, 3) float64
    forces                   (N, 3) float64
  constraint 1/
    ...
\end{lstlisting}

\subsubsection{TorsionScanRelax Subset}

The CSV metadata for the TorsionScanRelax subset includes the following columns:
\begin{itemize}
  \item \texttt{uuid}
  \item \texttt{mapped\_nonisomeric\_smiles}
  \item \texttt{mapped\_isomeric\_smiles}
  \item \texttt{torsion\_indices}
  \item \texttt{h5\_file}
  \item \texttt{num\_constraints}
  \item \texttt{num\_total\_steps}
\end{itemize}
The HDF5 internal structure is defined as follows:
\begin{lstlisting}[basicstyle=\ttfamily\small, frame=lines]
/<uuid>/
  mapped_nonisomeric_smiles  (utf-8 string object)
  mapped_isomeric_smiles     (utf-8 string object)
  atomic_numbers             (N, 1) int32
  torsion_atom_indices       (4,) int32        # 0-based [i,j,k,l]
  constraint 0/
    energy                   (M,) float64      # M is the number of steps
    coords                   (M, N, 3) float64
    forces                   (M, N, 3) float64
  constraint 1/
    ...
\end{lstlisting}

\subsubsection{MBIS Subset}

The CSV metadata for the MBIS subset includes the following columns:
\begin{itemize}
  \item \texttt{uuid}
  \item \texttt{mapped\_nonisomeric\_smiles}
  \item \texttt{mapped\_isomeric\_smiles}
  \item \texttt{h5\_file}
\end{itemize}
The HDF5 internal structure is defined as follows:
\begin{lstlisting}[basicstyle=\ttfamily\small, frame=lines]
/<uuid>/
  mapped_nonisomeric_smiles  (utf-8 string object)
  mapped_isomeric_smiles     (utf-8 string object)
  atomic_numbers             (N, 1) int32
  coords                     (N, 3) float64
  mbis_info/
    atomic_volumes           (N, 1) float64
    atomic_charge            (N, 1) float64
    atomic_dipole            (N, 3) float64
    atomic_quadrupole        (N, 3, 3) float64
  parameters                 (M, 3) float64    # M MBIS Slater functions.
\end{lstlisting}
Each row in the \texttt{parameters} array contains the 0-based index of the parent atom, the charge population (amplitude) of the Slater function, and the inverse width (decay constant) of the function.

\subsection{Dataset Statistics}

\begin{table}[H]
  \centering
  \caption{
    Element counts at the molecule level across all subsets (excluding hydrogen). Values in parentheses indicate the percentage of molecules containing the element.
  }
  \label{tab:element_molecule}
  \resizebox{\textwidth}{!}{
  \begin{tabular}{lrrrrrrr}
    \toprule
    Element & Hessian & HessianRelax & TorsionScan & TorsionScan (unique) & TorsionScanRelax & TorsionScanRelax (unique) & MBIS \\
    \midrule
    C & 3100300 (99.93\%) & 4809554 (99.95\%) & 4192260 (99.99\%) & 2436502 (99.98\%) & 4914033 (99.99\%) & 3089945 (99.98\%) & 3079875 (99.93\%) \\
    N & 2511706 (80.96\%) & 4210445 (87.50\%) & 3600678 (85.88\%) & 2062283 (84.62\%) & 4336181 (88.23\%) & 2709037 (87.66\%) & 2488345 (80.73\%) \\
    O & 2355077 (75.91\%) & 3490226 (72.54\%) & 3409587 (81.32\%) & 1897151 (77.85\%) & 3834034 (78.01\%) & 2293603 (74.21\%) & 2337180 (75.83\%) \\
    S & 635415 (20.48\%) & 948882 (19.72\%) & 1032489 (24.63\%) & 557524 (22.88\%) & 1174183 (23.89\%) & 689218 (22.30\%) & 624439 (20.26\%) \\
    F & 405535 (13.07\%) & 563136 (11.70\%) & 602375 (14.37\%) & 342030 (14.03\%) & 684840 (13.93\%) & 410136 (13.27\%) & 401583 (13.03\%) \\
    Cl & 403989 (13.02\%) & 605830 (12.59\%) & 558673 (13.32\%) & 328770 (13.49\%) & 655559 (13.34\%) & 412561 (13.35\%) & 402773 (13.07\%) \\
    Br & 178050 (5.74\%) & 238083 (4.95\%) & 230139 (5.49\%) & 138115 (5.67\%) & 263730 (5.37\%) & 166190 (5.38\%) & 149842 (4.86\%) \\
    Si & 282990 (9.12\%) & 264793 (5.50\%) & 82666 (1.97\%) & 82666 (3.39\%) & 92786 (1.89\%) & 92786 (3.00\%) & 283165 (9.19\%) \\
    B & 69851 (2.25\%) & 74357 (1.55\%) & 55657 (1.33\%) & 55657 (2.28\%) & 60870 (1.24\%) & 60870 (1.97\%) & 78088 (2.53\%) \\
    P & 29970 (0.97\%) & 35976 (0.75\%) & 43440 (1.04\%) & 24117 (0.99\%) & 42646 (0.87\%) & 26164 (0.85\%) & 30032 (0.97\%) \\
    I & 25923 (0.84\%) & 31976 (0.66\%) & 27561 (0.66\%) & 18815 (0.77\%) & 30449 (0.62\%) & 21316 (0.69\%) & 25909 (0.84\%) \\
    \bottomrule
  \end{tabular}
  }
\end{table}

\begin{table}[H]
  \centering
  \caption{
    Element counts at the atom level across all subsets (excluding hydrogen). Values in parentheses indicate the percentage of atoms of the specific element.
  }
  \label{tab:element_atom}
  \resizebox{\textwidth}{!}{
  \begin{tabular}{lrrrrrrr}
    \toprule
    Element & Hessian & HessianRelax & TorsionScan & TorsionScan (unique) & TorsionScanRelax & TorsionScanRelax (unique) & MBIS \\
    \midrule
    C & 37016882 (73.85\%) & 58147021 (72.68\%) & 51473740 (73.10\%) & 29332995 (73.55\%) & 59226366 (72.39\%) & 36383019 (72.49\%) & 36772795 (73.90\%) \\
    N & 5800250 (11.57\%) & 11766955 (14.71\%) & 8462295 (12.02\%) & 4811522 (12.06\%) & 11061709 (13.52\%) & 7053590 (14.05\%) & 5746390 (11.55\%) \\
    O & 4733029 (9.44\%) & 6646077 (8.31\%) & 7145702 (10.15\%) & 3791825 (9.51\%) & 7752802 (9.48\%) & 4413332 (8.79\%) & 4708892 (9.46\%) \\
    S & 717426 (1.43\%) & 1064014 (1.33\%) & 1184010 (1.68\%) & 631170 (1.58\%) & 1336913 (1.63\%) & 776032 (1.55\%) & 704549 (1.42\%) \\
    F & 669673 (1.34\%) & 889364 (1.11\%) & 988095 (1.40\%) & 560470 (1.41\%) & 1103008 (1.35\%) & 656452 (1.31\%) & 664497 (1.34\%) \\
    Cl & 510049 (1.02\%) & 750859 (0.94\%) & 683969 (0.97\%) & 409638 (1.03\%) & 800743 (0.98\%) & 511149 (1.02\%) & 508978 (1.02\%) \\
    Br & 195447 (0.39\%) & 260264 (0.33\%) & 250788 (0.36\%) & 150873 (0.38\%) & 287226 (0.35\%) & 181422 (0.36\%) & 166604 (0.33\%) \\
    Si & 345578 (0.69\%) & 323436 (0.40\%) & 92182 (0.13\%) & 92182 (0.23\%) & 103533 (0.13\%) & 103533 (0.21\%) & 345802 (0.69\%) \\
    B & 72740 (0.15\%) & 77350 (0.10\%) & 56092 (0.08\%) & 56092 (0.14\%) & 61369 (0.08\%) & 61369 (0.12\%) & 81297 (0.16\%) \\
    P & 32904 (0.07\%) & 39472 (0.05\%) & 47838 (0.07\%) & 26299 (0.07\%) & 46343 (0.06\%) & 28203 (0.06\%) & 32984 (0.07\%) \\
    I & 28009 (0.06\%) & 34404 (0.04\%) & 30331 (0.04\%) & 20404 (0.05\%) & 33364 (0.04\%) & 23066 (0.05\%) & 27991 (0.06\%) \\
    \bottomrule
  \end{tabular}
  }
\end{table}

\end{document}